\documentclass[conference]{IEEEtran}

\usepackage{floatflt}

\usepackage{color}
\usepackage{graphicx}
\usepackage{amsfonts}
\usepackage{amsmath}
\usepackage{graphicx}
\usepackage{cite}
\usepackage{epsfig}
\usepackage{latexsym}
\usepackage{subfigure}
\usepackage{epstopdf}
\usepackage{verbatim}
\usepackage{units}
\usepackage{amsthm}
\usepackage[longend,ruled]{algorithm2e}
\usepackage{booktabs}

\begin{document}
\title{RF Fingerprinting and Deep Learning Assisted UE Positioning in 5G}
\author{\IEEEauthorblockN{M. Majid Butt\IEEEauthorrefmark{1},~Anil Rao\IEEEauthorrefmark{2},~Daejung Yoon\IEEEauthorrefmark{1}}
\IEEEauthorblockA{\IEEEauthorrefmark{1}Nokia Bell Labs, Paris-Saclay, France\\
\IEEEauthorblockA{\IEEEauthorrefmark{2}Nokia Bell Labs, Naperville, USA\\
Email: \{majid.butt, anil.rao, daejung.yoon\}@nokia-bell-labs.com}}
}
\maketitle
\begin{abstract}
In this work, we investigate user equipment (UE) positioning assisted by deep learning (DL) in 5G and beyond networks. As compared to state of the art positioning algorithms used in today's networks, radio signal fingerprinting and machine learning (ML) assisted positioning requires smaller additional feedback overhead; and the positioning estimates are made directly inside the radio access network (RAN), thereby assisting in radio resource management. The conventional positioning algorithms will be used as back-up for the environments with high variability in conditions; but ML-assisted positioning serves as more efficient and simpler technique to provide better or similar positioning accuracy. In this regard, we study ML-assisted positioning methods and evaluate their performance using system level simulations for an outdoor scenario in Lincoln park Chicago. The study is based on the use of raytracing tools, a 3GPP 5G NR compliant system level simulator and DL framework to estimate positioning accuracy of the UE. The use of raytracing tool and system level simulator helps avoid expensive drive test measurements in practical scenarios. Our proposed mechanism is a first step towards more proactive mobility management in future networks. We evaluate and compare performance of various DL models and show mean positioning error in the range of 1-1.5m for the best DL configuration with appropriate system feature-modeling.
\end{abstract}

\section{Introduction}
\subsection{Motivation and Background for UE Positioning}
5G technologies introduce a new paradigm to connectivity and mobility by using accurate  location information. Positioning and localization studies have been conducted in various research areas with various systems and devices for many decades such as global positioning systems, industrial internet of things, autonomous driving, indoor maps and so on \cite{positioning_NR,positioning_LTE}. Localization applications are found from small indoor area to wide landscapes of satellite coverage. As mobile handsets are widespread, users start utilizing localization services through cellular networks. In the 5G era, various types of even more devices appear such as sensor devices, robots and vehicles, all operating based on highly accurate location information.

Positioning accuracy is one of the most important factors to evaluate localization system performances. Federal Communications Commission (FCC) in the United States has set the E911 regulatory performance requirements \cite{FCCE911} as well as further accuracy requirements defined tightly by commercial systems \cite{positioning_NR}. FCC E911 requires horizontal error less than 50m and vertical error less than 5m for $80\%$ of UEs \cite{FCCE911}. In commercial systems like 3GPP, requirements are tighter than E911 ones; 3GPP NR positioning techniques and requirement are under discussion with horizontal positioning error less 10m and vertical error less than 3m \cite{positioning_NR}.
In addition, measurement cost and efficiency of localization service are getting more important, since tiny devices appearing in specific use cases require localization services in a small area.

In order to achieve such goals of high accuracy and measurement efficiency, various positioning methods have been deeply studied. In localization systems, usually a positioning method is defined depending on a type of measurement matrices. Timing-based techniques take advantage of knowledge of time-differences due to the speed of RF signals for distance calculation between a transmitter and receiver. The measurement matrix is estimation of the time of arrival (TOA) of signals. Angle-based techniques take advantage of transmit beamforming of signals and differences of phase across receive antenna elements to determine the angle between a transmitter and receiver.

Machine learning algorithms are making contributions to improve both measurement accuracy and efficiency in hybrid techniques. ML assisted positioning is based on fingerprinting locations with radio signal data and then UE position is inferred online.


%

\begin{figure*}
\centering
  	\includegraphics[width=4.2in]{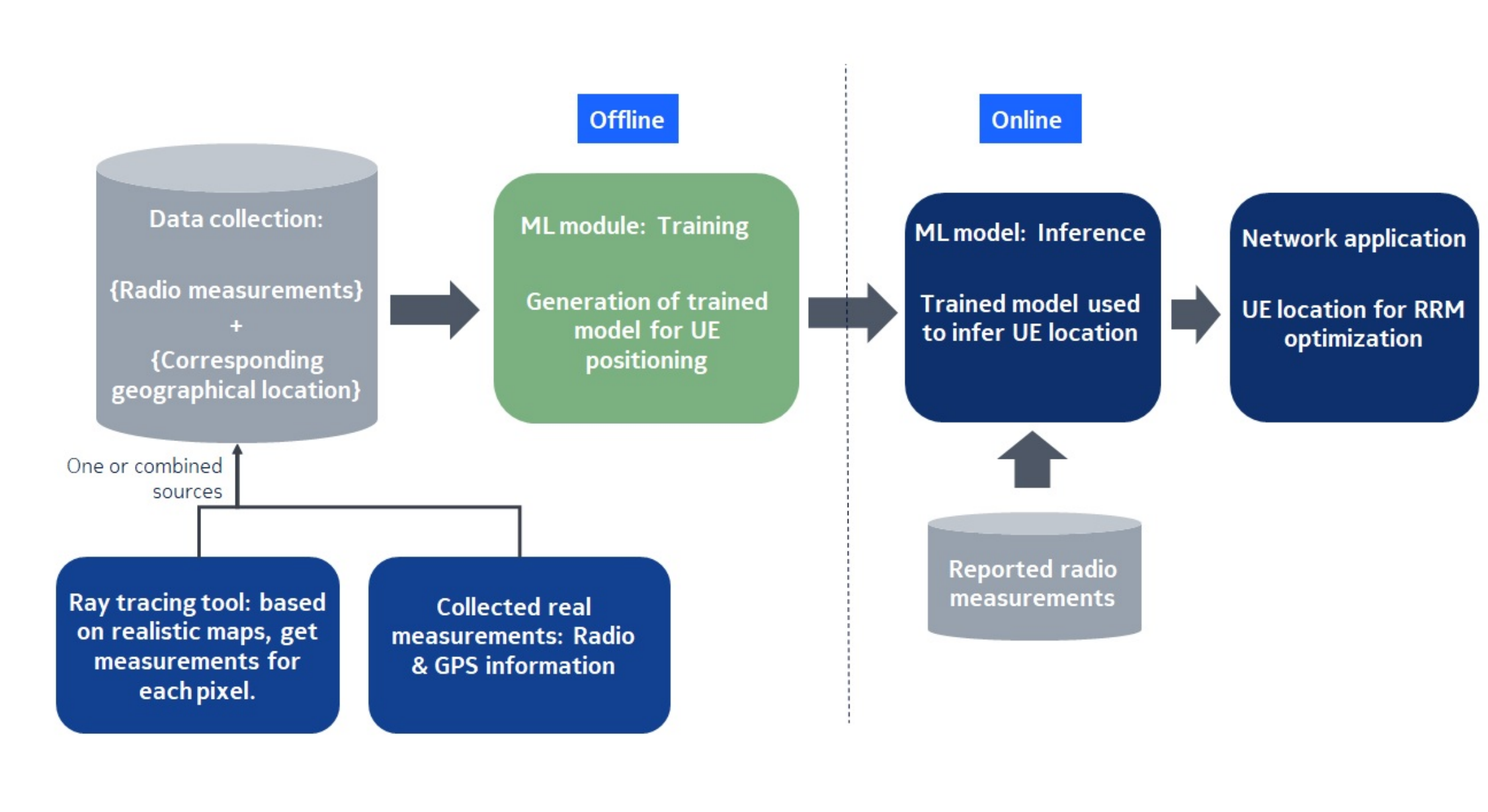}
   \caption{Schematic diagram for our proposed framework. Offline data preprocessing and ML training is performed while online part performs inference.}
	\label{fig:model}
\end{figure*}
\subsection{Standardization on Positioning}
3GPP positioning feature has been supported in Rel-9 LTE networks; several RF based positioning methods have been introduced, and advanced techniques are under investigation for NR in Rel-16. Recent 3GPP NR positioning study conclusions have been stated in \cite{positioning_3GPP}. There has been discussion of many different possible positioning solutions in the study item, conclusions have been made that the 3GPP NR systems should support solutions of observed time difference of arrival (OTDOA), uplink time difference of arrival (UTDOA), angle information, Multi-Cell Round Trip Time (Multi-RTT), and enhanced cell-ID (E-CID) in NR Rel-16 \cite{qualcom2014}. Associated with the solutions, 3GPP is under discussion to define network measurements through DL or UL radio link. OTDOA requires UEs to measure DL-TDOA. UEs measure TDOA based on DL positioning reference signals from the neighbor and serving cells. A location server collects TDOA measurements and compute a UE location. UTDOA requires gNBs to measure ToA based on UL positioning reference signal.
Also, angle measurement is an important information for positioning.

One of the major changes in NR compared with legacy cellular networks is the addition of the millimeter wave (mmWave) frequency band.
NR also provides support for beam-based operations. During the beam search operation, angle can be obtained easily in terms of downlink angle of departure (DL-AoD) and uplink angle of arrival (UL-AoA). While UEs and gNBs search for the best beam for radio link, DL and UL reference signal received  power (RSRP) is measured to evaluate signal quality of a beam.
Received power measurement has been considered for positioning finger print in wireless systems \cite{pattern_matching_3GPP}. In LTE system, E-CID method is widely used for positioning based on the received signaling power. The location server constructs a fingerprint for measurement-to-location mapping, but it needs to involve measurements from many cells for high accuracy.


\subsection{State of the Art and Contributions}
As the NR physical layer design inherently supports beam-based RSRP measurement reports from the UE for multiple beams, we study how well ML-based approaches work to determine the UE position directly from these measurements. A significant advantage is that when it comes to estimating the UE position during the ML inference phase, the only inputs needed are measurements the UE is providing to its serving cell, hence the positioning estimation can be done using an ML inference engine in the baseband of the UE's serving cell, avoiding the need for the location server framework which lives outside the radio access network and therefore, efficiently supporting a single-cell based positioning method.

ML techniques have been used to solve various problems in communication systems. References \cite{Simeone:TCN,Shea:TCN_2017,DL_survey:2019} survey some exciting works on use of ML algorithms in the field of wireless communication and networking. Some recent works apply ML techniques for localization problems. In \cite{Xiao} autoencoder-based indoor localization method is studied that provides high-performance 3-D positioning in large indoor places for bluetooth low energy networks. The authors in \cite{Prasad_TWC:2018} propose a machine learning approach based on Gaussian process regression to position users in a distributed massive multiple-input multiple-output (MIMO) system with the uplink received signal strength (RSS) data.
The simulation results show that the proposed method improves performance in massive MIMO scenarios. In \cite{Ye_IEEEAceess_2017}, neural networks are used on input of fingerprint data obtained from channel characteristics and geographical locations and median error of 6 and 75 meters is reported for indoor and outdoor scenarios, respectively. Similar approaches for UE positioning using neural networks (NNs) are investigated in \cite{Wu_electronics:2019,Lee_ICASSP:2018} as well. However, most of these works provide very high positioning error compared to our results and feature vectors are processed differently.

We provide a complete system level study starting from generating raytracing data, using system level simulator to generate RSRP fingerprinting for geographical $(X,Y)$ coordinates; and finally applying ML techniques to train and validate UE positioning accuracy.
We study various input features that contribute to improve positioning accuracy and conclude that a combination of RSRP values from the serving and neighboring cell beams provides the best results. Then, we evaluate the performance of network level and cell-specific training topologies and quantitatively show that cell-specific training outperforms network level training. For comparison sake, we evaluate decision tree regression ML technique to evaluate the performance and observe reasonably close performance to DL.

Rest of this paper is organized as follows. System set up for our experimentation is described in detail in Section \ref{sec:system_setup}. We discuss the machine learning approaches used in this work in Section \ref{sec:ML methods}. Performance of the ML approach is evaluated and discussed in Section \ref{sec:perfromance}, while Section \ref{sec:conclusions} concludes with the main contributions of the paper.

\section{System Setup}
\label{sec:system_setup}

In this section, we describe the system set up for our study. As shown in Fig. \ref{fig:model}, the system set up is divided into following main components:

\begin{enumerate}
  \item {\textbf{Raytracing}: The raytracing software WinProp is used to generate raytracing data as a function of $(X,Y)$ coordinates, and full 3-D multipath ray data is generated for each point. In this case, an outdoor urban database and the Intelligent Ray Tracing (IRT) algorithm were configured.}
  \item {\textbf{System Level Simulation}: Based on the raytracing data generated for channel modelling, system level simulation of the 3GPP 5G NR air interface including beamforming is run to produce beam RSRP values for each UE. The beam RSRP values serve as fingerprinting for the under-study area based on raytracing data.}
  \item{\textbf{Offline ML Traning}: ML techniques are then applied on the RSRP data received for an offline training.}
  \item{\textbf{Online ML Inference}: Finally, trained ML model is evaluated to determine position of the UEs. \textit{Euclidean distance} between the estimated and actually UE position serves as the accuracy measure for the estimated UE position.}
\end{enumerate}
The last block in Fig. \ref{fig:model} points toward possible use cases for application of accurately estimated positioning data.

\begin{figure}
\centering
  	\includegraphics[width=2.7in]{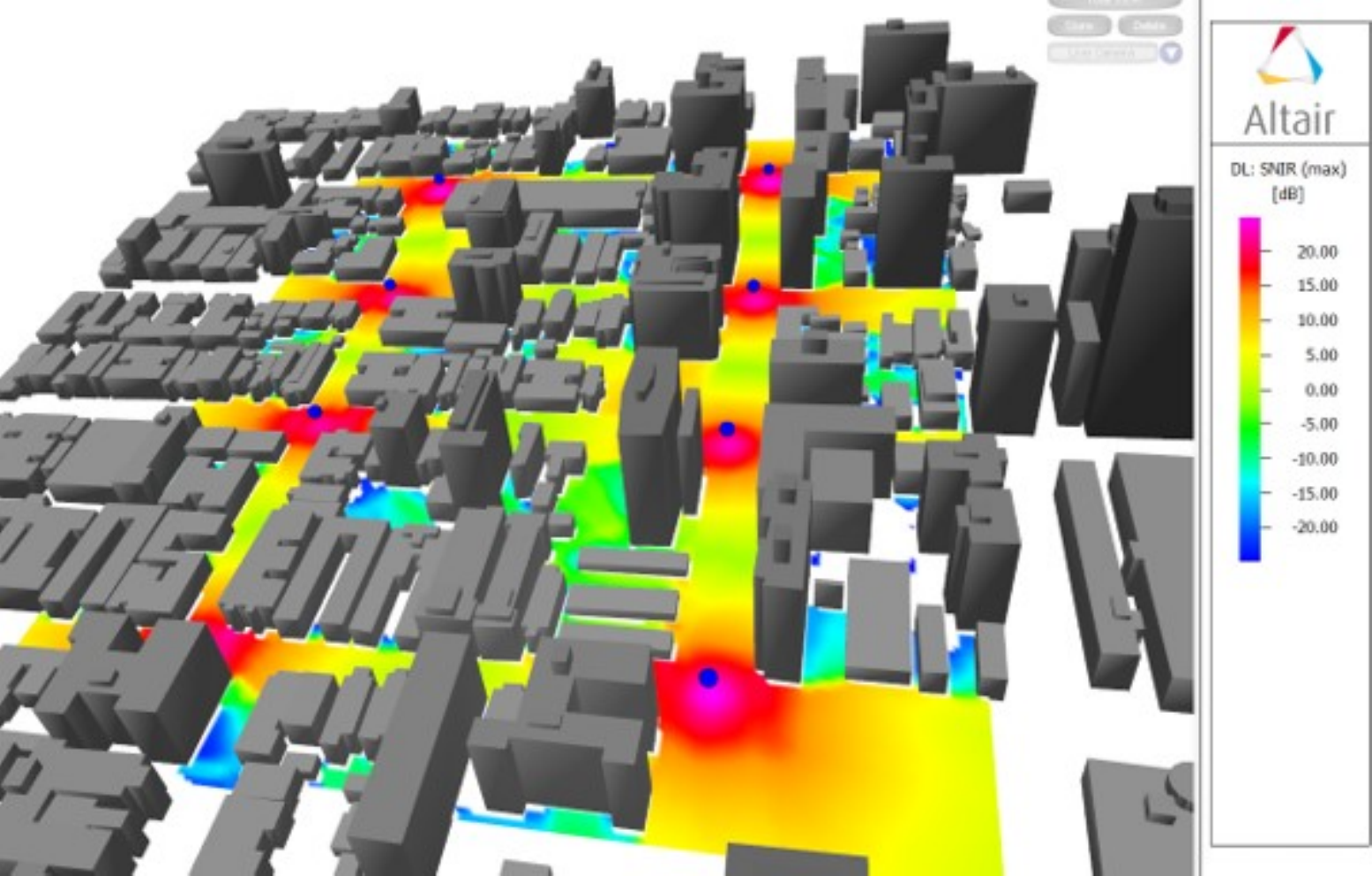}
  \vspace{-0.2cm}
   \caption{8 sites are placed at 8 corners of Lincoln Park, Chicago for the purpose of raytracing data generation.}
	\label{fig:Raytracing}
\end{figure}

Raytracing is performed for a section of Lincoln Park, Chicago. Eight sites are placed at each street corner in as shown in Fig. \ref{fig:Raytracing}. In this study, we have confined ourselves to study of positioning accuracy for the UEs that have an Line of sight (LoS) connection to their serving cell. We observed that positioning accuracy using this ML-assisted method in the mmWave propagation environment is severely affected when LoS is not available to the UE. However, this may not be a big issue in future networks where very dense mmWave deployments are planned and LoS non-availability chances will not be that high. In our study, $62\%$ of the UEs have LoS available and we consider only those users for ML training and testing. For ML training purpose, we only consider RSRP feature for the beams in current time slot, i.e. a snapshot based approach is employed. The methodology used in this study provides an efficient option to evaluate feasibility of this approach compared to drive testing.

We use two parameters, mean and standard deviation denoted by $\bar{E}$ and $\sigma$ respectively, of the Euclidean distance between the estimated training/test UE position and the actual UE position and define them as,
\begin{eqnarray}
  \bar{E} &=& \frac{\sum_{n=1}^N x_n}{N} \\
  \sigma &=& \sqrt{\frac{1}{N}{\sum_{n=1}^N (x_n-\bar{E})^2}}
\end{eqnarray}
where $x_n$ is value for UE Euclidean distance for $n^{th}$ sample.

\section{Machine Learning Techniques}
\label{sec:ML methods}
Deep Learning and decision tree ML techniques have been used to solve regression problem for accurate UE position estimation. We briefly explain both techniques in the following.

\subsection{Summary of Deep Learning Parametrization}
Deep learning is one of the powerful ML techniques to solve classification/regression problems. Deep feedforward networks are advanced neural networks (NNs) with one or more hidden layers of neurons between input and output layer. In \emph{feedforward NNs}, information flow from input $x$ to output $y$ through intermediate layers and there is no feedback, contrast to Recurrent NNs where information is fed back as well.


A feedforward deep NN (DNN) has an input layer, which is connected with output layer through hidden layers. The hidden layers represent \emph{depth} of the network, and 'deep learning' term arises from this. The number of neurons in each layer represent \emph{width} of an NN where neuron is a computational unit that represents a vector-to-scalar mapping. An activation function, such as 'tanh' and 'Rectified Linear Units (ReLU)', is used to compute element (neuron) wise non-linear transformation of learned parameters 'weights' and 'bias', which is transferred to next layer. Goal of a DNN is to learn the function that maps input $x$ to output $y$. Input vector $x$ is termed as \emph{feature vector} in DL literature. Architecture of a neural network defines how many layers a NN has, how many neurons per layer are required and how layers are connected. This is a stochastic process and termed as 'hyperparameter' optimization.\footnote{Hyperparameter optimization is one of the main tasks in ML and include optimization of several parameters by experimentation.}

A deep NN is first \emph{trained} on large number of examples/measurements provided where solution $y$ is provided for the input $x$, called supervised learning. 'Loss' is defined as the difference between the computed output and actual output parameterized by some function, such as minimum squared error (MSE), entropy function, etc.
Based on the training data, DNN tries to learn mapping function that minimizes 'loss' as training progresses. Once the function is computed, it is 'tested' against the data not part of training to validate the computed DNN model. Test error is then computed by appropriate metric to compute accuracy of the trained DNN.

In ML, 'overfitting' and 'underfitting' are two important concepts. 'Overfitting' represents the situation when DNN is over-trained and it becomes hard for the computed model to generalize for the unknown testing examples.
To overcome this problem, 'early stopping' and 'regularization' is used. On the other side, 'underfitting' represents the situation when both training and testing error are large.
Analysis of training and testing error provides good indication of accuracy of DNN model and helps designer take right actions to improve the model. Collecting more data and/or hyperparameter optimization are the well known solutions.

This section just provided a brief introduction to DNN. For further details on the terms defined and their explanation, the interested reader is referred to \cite{Goodfellow-et-al-2016}.
\subsection{Decision-Tree Regression}
For a comparison purpose, we use Decision Tree regressor instead of DL framework. Decision Trees is a supervised learning method used for classification and regression for the problems where feature vectors have 'tree like' structure. The goal is to train a model that predicts the value of a target variable by learning decision rules inferred from the data features. Decision trees are simple to implement and results can be easily visualized and analyzed. As we will see later, our feature vector has a 'tree like' structure and suits very well to regression problem on hand.
\section{Performance Evaluation}
\label{sec:perfromance}

\begin{table}
\begin{center}
\footnotesize
\caption{Raytracing Parameters}
\vspace{-0.3cm}
\begin{tabular}{lr}
\toprule
Parameter & Value\\
\midrule
Number of sites							& 8\\
Access Point height & 10m\\
Inter-site distance& 110m vertical, 200m horizontal\\
Carrier frequency & 28 GHz\\
Resolution of data points & 1m (123,768 locations sampled)\\
\bottomrule
\end{tabular}

\label{tab:winprop}
\end{center}
\end{table}

\begin{table}
\begin{center}
\footnotesize
\caption{System Level Simulation Parameters}
\vspace{-0.3cm}
\begin{tabular}{lr}
\toprule
Parameter & Value\\
\midrule

Number of SSB beams &32\\
Number of gNBs 	 					& 8 having 3 sectors each\\
DL Tx Power per sector & 30 dBm\\
Element Azimuth Beamwidth&65 degrees\\
Element Elevation Beamwidth&65 degrees\\
Element gain & 8dBi\\
Front2Back & 30dB\\
Mechanical downtilt & 5 degrees\\
\bottomrule
\end{tabular}

\label{tab:system_level}
\end{center}
\vspace{-0.4cm}
\end{table}

\begin{figure*}
\centering
  \subfigure[Example 1]
  	{\includegraphics[width=3.0in]{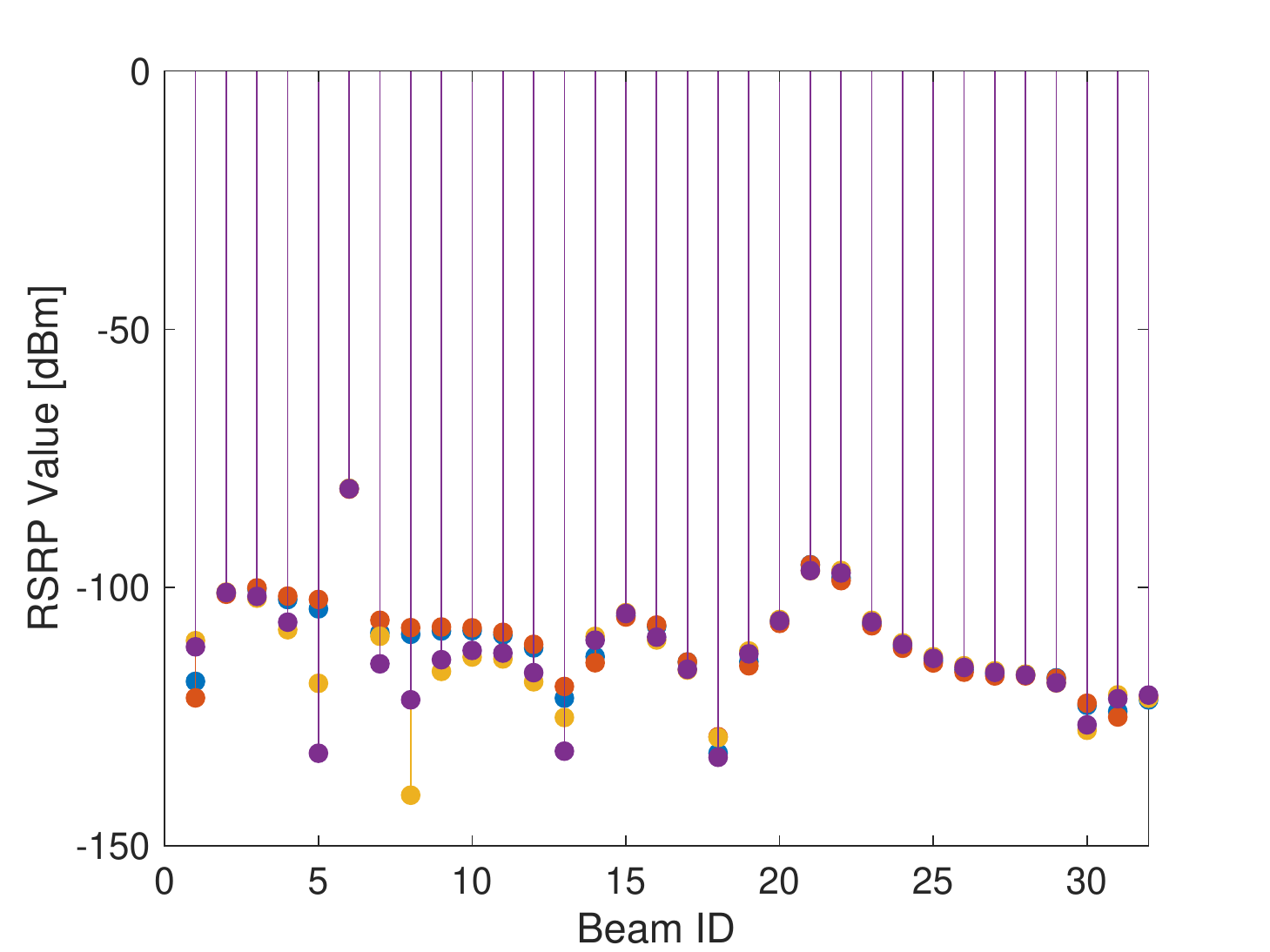}
  \label{fig:nonoverlape}}
 \subfigure[Example 2]
  	{\includegraphics[width=3.0in]{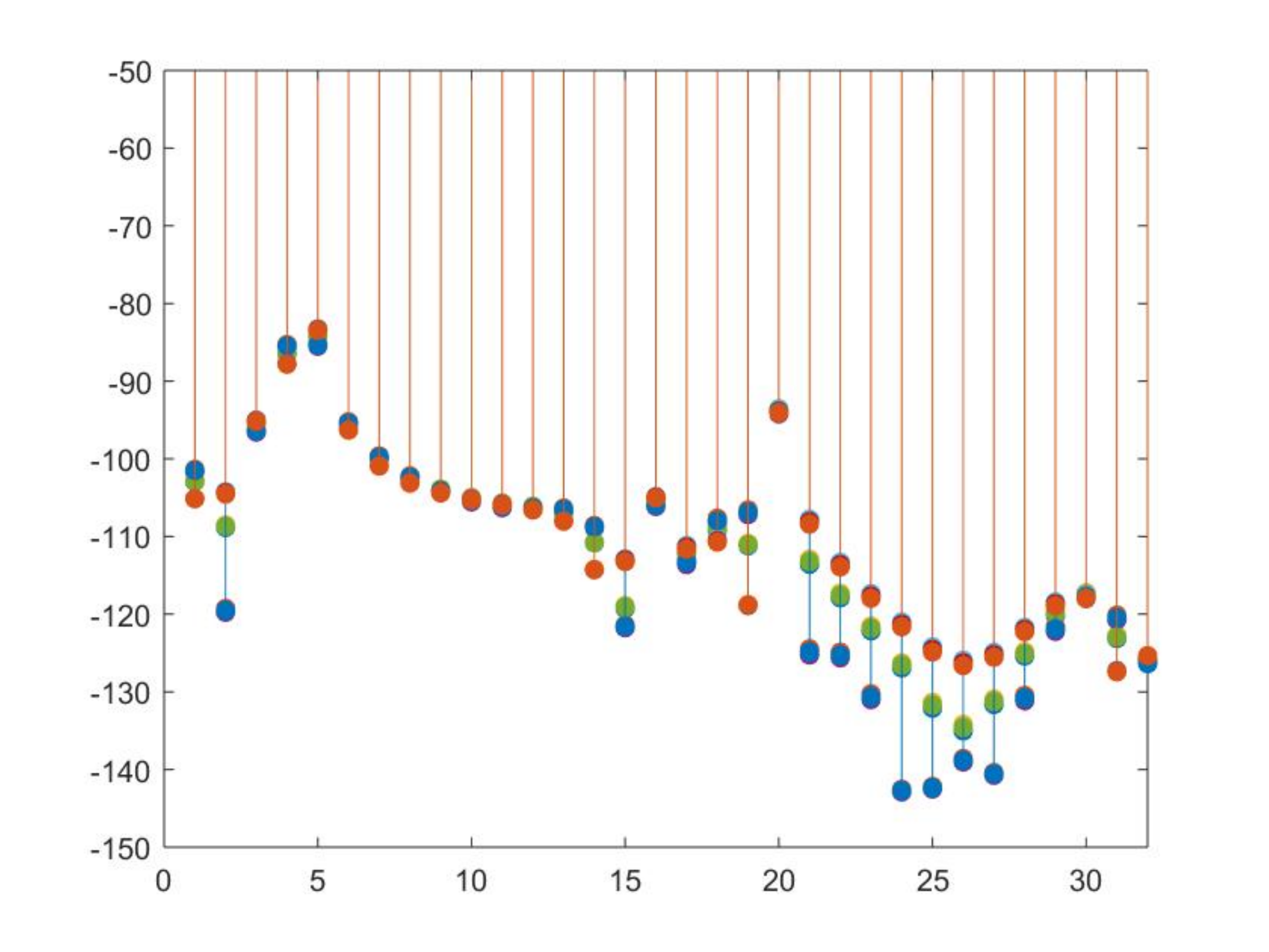}
  \label{fig:overlape}}
   \caption{Pattern for RSRP values for the 32 beams received at the UEs in close proximity.}
	\label{fig:RSRP}
\end{figure*}
Parameters used for raytracing in Winprop software and system level simulations are summarized in Table \ref{tab:winprop} and \ref{tab:system_level}, respectively. The antenna panel used for system level simulations was a $16\times16$ cross-pol array, formed of 16 azimuthal beams and 3 elevation beams.
For evaluation of DL module, the parametric settings are explained as follows. 90$\%$ of the labeled data is used for training and $10\%$ is set aside for validation. Other parameters include, Batch size = 32, 'tanh' activation function for input and hidden layers, linear activation function for output layer, 'adam' optimizer, loss metric = 'MSE'. We used Keras API with TensorFlow backend. 'Early Stopping' based on the 'loss' in the training data is set to prevent overfitting while maximum number of epochs is set to 500. Data is normalized using mean and standard deviation normalization.

\subsection{Network-Level DL Training}
\label{sec:network_level}
We first study the case where ML training is performed over all the data available from 24 sites at one positioning server. Current 3GPP Release 15 standard supports UE reporting for up to 4 strongest beams from the serving cell. In Table \ref{tab:comparison bb}, we show training and testing accuracy for the case when RSRP values for 4 strongest beams from the serving cell, corresponding beam IDs and cell ID is the input for the input layer of the DNN. This will serve as a baseline. For the 9 input features, we train our DNN. We use only one hidden layer to get some idea on suitability of features and vary the number of neurons for each experiment\footnote{Multilayer NNs improve accuracy as we show later, however we confine ourselves to single hidden layer in the beginning to get more insight on the other parameters influencing our model.}. Estimation accuracy improves with increasing number of neurons and we get mean test accuracy of approximately 5m for very large number of neurons.

\begin{table}
\caption{Positioning Accuracy for 4 Strongest beams from the Serving Cell}
\label{tab:comparison bb}
\begin{center}
\begin{tabular}{|c|c|c|c|c|}
\hline
  No. &Training $\bar{E}$  &Training $\sigma$  &Test $\bar{E}$ &Test $\sigma$ \\
  Neurons& (m)& (m)&  (m)&(m)\\
  \hline
  576&5.4&4.3&5.7&5.1\\
  1152&4.3&3.4&{\large\textcircled{\footnotesize 4.9}}&4.4\\
  2304&4.1&3.1&5.1&4.7\\
  \hline
\end{tabular}
\end{center}
\end{table}

In Fig. \ref{fig:RSRP}, we investigate the main hurdle in improving positioning accuracy based on RSRP values of the beams from the serving cell alone. The figures are plotted by choosing a UE randomly and then finding the UEs in its close proximity. We observe high correlation in serving cell beam IDs and their respective RSRP values for the UEs, which makes it harder to improve positioning accuracy of the UEs in close proximity. Fig. \ref{fig:nonoverlape} and Fig. \ref{fig:overlape} show 2 such examples where strongest beam IDs as well as their RSRPs are almost overlapping. After some further experimentation with various number of strongest beams from the serving cell with DNN (not reported in this paper), we observe a negligible difference in positioning accuracy by using 3 instead of 4 beams from the neighboring cell. As UE reporting is costly operation, we use RSRP values for at most 3 beams from the serving cell for further investigation in this sequel.

\begin{table}
\caption{Positioning Accuracy for RSRP data from both Serving and Neighboring Cells}
\label{tab:serving and neighboring}
\begin{center}
\begin{tabular}{|c|c|c|c|c|c|}
\hline
  No. beams from &Training $\bar{E}$  &Training $\sigma$  &Test $\bar{E}$ &Test $\sigma$ \\
  neighbor cells& (m)& (m)&  (m)&(m)\\
  \hline
  0&4.3&3.6&4.8&4.7\\
  1&2.8&2.6&3.6&4.8\\
  2&2.2&1.5&{\large\textcircled{\footnotesize 3.0}}&3.4\\
  3&2.1&1.5&3.1&4.9\\
  \hline
\end{tabular}
\end{center}
\end{table}

Based on our investigation, it is necessary to change our feature vector for the input layer. We propose to make use of RSRP values of the beams not only from the serving cell but from the neighboring cells as well. In Table \ref{tab:serving and neighboring}, we show position accuracy results by fixing number of strongest beams from the serving cell to 3 and varying number of beams from the neighbor cells. For example, using 2 strongest beams from 2 neighbor cells alongwith their beam and cell IDs give us 6 more features in addition to 7 features from 3 beam IDs, their respective RSRP values and cell ID. From Table \ref{tab:serving and neighboring}, we observe that mean accuracy of 3m is achieved by using data from 2 strongest beams from the neighboring cells in addition to 3 serving cell beams. This is considerable improvement over the case using data from beams only from the serving cell. This performance gain is achieved by adding more features and date for the DNN model training.


\subsection{Cell-Specific DL Training}
\label{sect:Cell-specific}
To further improve the performance, we investigate the effect of change in location of training. In Section \ref{sec:network_level}, the DNN training was performed with all the data from all the sites at one location server. In this section, we study positioning accuracy by training individual cell-specific DNN model for each cell. This results in removing the serving cell ID feature from the feature vector as it is known a-priori, called dimensionality reduction in ML literature. We preprocess data generated by the system level simulator to differentiate data for various cells. As a result, size of both training and test data may vary from cell-to-cell. For example, we have data for more than $10\times 10^3$ positions for cell number 14, while it is about $700$ for cell number 1. From Table \ref{tab:winprop}, we know that we had more than $122\times 10^3$ location samples for the cell level training gathered over all 24 sites. Lack of data could pose a major challenge in training of DNNs in Cell-Specific training.

\begin{table}
\caption{Cell-Specific, Cell ID 14, Single Hidden Layer: Positioning Accuracy for data from both Serving and Neighboring Cells}
\label{tab:cell_specific 14}
\begin{center}
\begin{tabular}{|c|c|c|c|c|c|}
\hline
  No. beams from &Training $\bar{E}$  &Training $\sigma$  &Test $\bar{E}$ &Test $\sigma$ \\
  neighbor cells& (m)& (m)&  (m)&(m)\\
  \hline
  0&2.9&3.0&3.3&3.1\\
  1&2.3&2.0&2.8&5.1\\
  2&1.7&1.2&{\large\textcircled{\footnotesize 2.0}}&2.1\\
  3&1.4&1.1&1.9&1.7\\
  \hline
\end{tabular}
\end{center}
\end{table}

In Table \ref{tab:cell_specific 14}, we study DNN with a single hidden layer and cell specific training for cell number 14. As previously observed, best feature vector uses data from 3 strongest beams from the serving cell. We again investigate accuracy by varying number of beams from the neighboring cells for cell-specific training. Again, positioning accuracy improves by using data from more beams from the neighboring cells. In this case, the best mean positioning accuracy is achieved for 3 neighboring cells; but difference is negligible as compared to 2 neighboring cell case. Looking at the cost of reporting information for this extra beam from the neighboring cell, we can safely assume that most of the performance gain is achieved with data from 3 serving and 2 neighboring cell beams. Remarkably, cell-specific training provides us mean positioning accuracy of 2m for 3 serving and 2 neighboring cell beams case which is considerably higher than the accuracy of 3m for the same configuration for network level training. This gain emerges from preprocessing of data and dimensionality reduction. After optimizing our network feature vector, we now look at further improving results by adding one more hidden layer to DNN. Table \ref{tab:cell-specific_2layers_14} shows that mean positioning error can be reduced to 1.4m by using 64 nodes in both hidden layers of DNN, a performance gain of 0.5m over the same parameters and single hidden layer case.

\begin{table}
\caption{Cell-Specific, Cell ID 14, 2 Hidden Layers: Positioning Accuracy for data from both Serving and Neighboring Cells}
\label{tab:cell-specific_2layers_14}
\begin{center}
\begin{tabular}{|c|c|c|c|c|c|}
\hline
  No. &Training $\bar{E}$  &Training $\sigma$  &Test $\bar{E}$ &Test $\sigma$ \\
  Neurons& (m)& (m)&  (m)&(m)\\
  \hline
  32&2.2&1.6&2.4&1.6\\
  64&1.1&0.7&{\large\textcircled{\footnotesize 1.4}}&1.2\\
  128&1.2&0.8&1.5&1.4\\
  \hline
\end{tabular}
\end{center}
\end{table}

\begin{figure}
\centering
  	\includegraphics[width=3.5in]{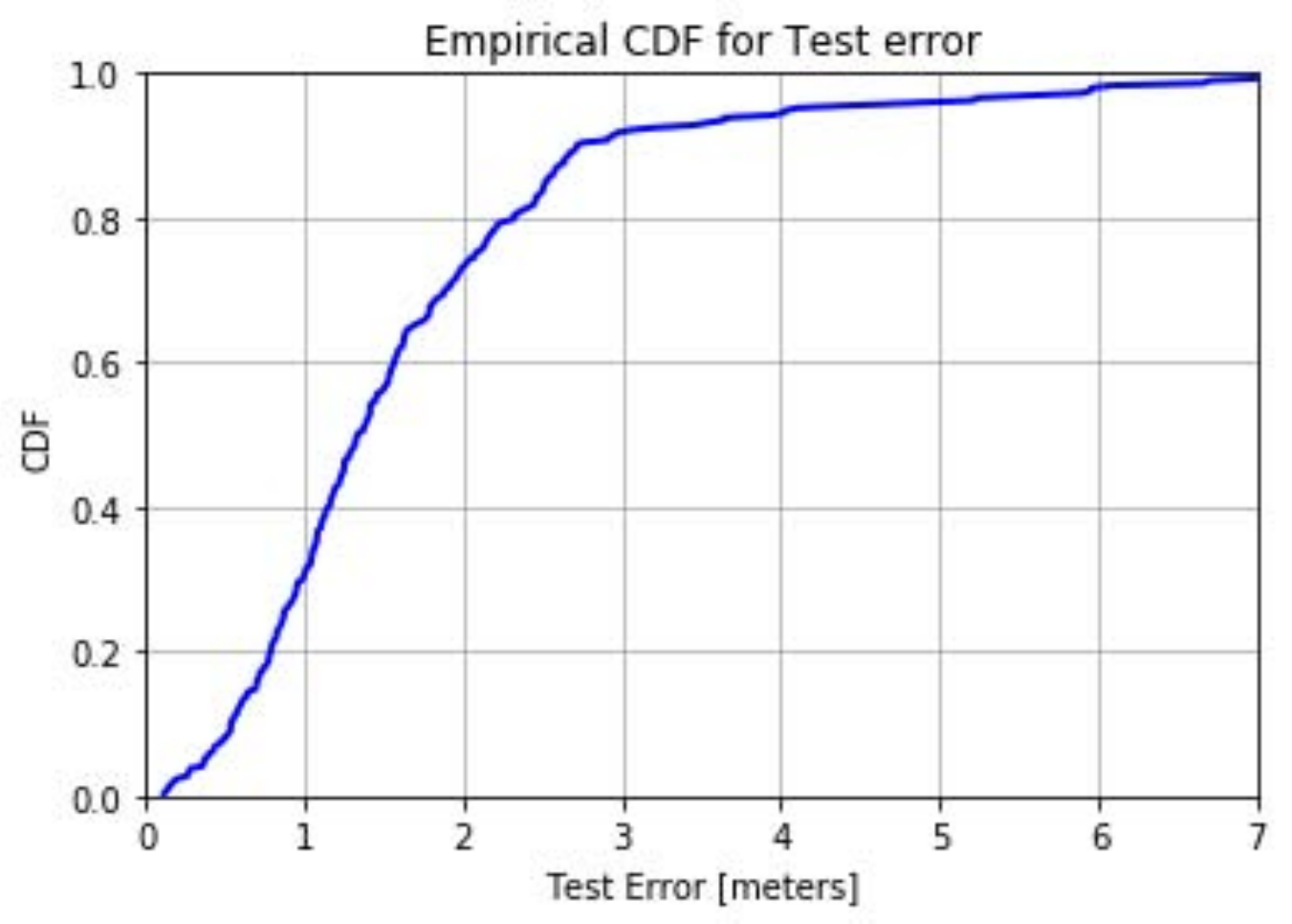}
  \vspace{-0.4cm}
   \caption{CDF of positioning accuracy for the 2 hidden layer, cell specific DL training for cell 14.}
	\label{fig:error_CDF}
\end{figure}

To elaborate further on the results, Fig. \ref{fig:error_CDF} shows cumulative distribution function (CDF) for this specific case that shows complete information of error for range of operation\footnote{CDF is the most commonly used criterion for reporting results in industry as it gives information about other percentile, e.g., $80\%$, $90\%$ percentile.}.

A clear improvement in position accuracy is observed over network level training. However, as we mentioned previously, lack of training data could be a potential bottleneck for cell-specific training. In Table \ref{tab:cell-specific_2layers_1}, we show results for the cell 1 with only 786 positioning examples. Note that, on average $62\%$ of these examples are LOS examples. Surprisingly, the results follow the results produced for cell 14 with relatively large data and produce identical positioning accuracy.

\begin{table}
\caption{Cell-Specific, Cell ID 1, 2 Hidden Layers: Positioning Accuracy for data from both Serving and Neighboring Cells}
\label{tab:cell-specific_2layers_1}
\begin{center}
\begin{tabular}{|c|c|c|c|c|c|}
\hline
  No. &Training $\bar{E}$  &Training $\sigma$  &Test $\bar{E}$ &Test $\sigma$ \\
  Neurons& (m)& (m)&  (m)&(m)\\
  \hline
  32&0.9&0.6&1.4&1.0\\
  64&1.1&0.8&{\large\textcircled{\footnotesize 1.4}}&0.9\\
  128&1.1&0.8&1.7&2.3\\
  \hline
\end{tabular}
\end{center}
\end{table}

\begin{figure}[t]
\centering
  	\includegraphics[width=3.5in]{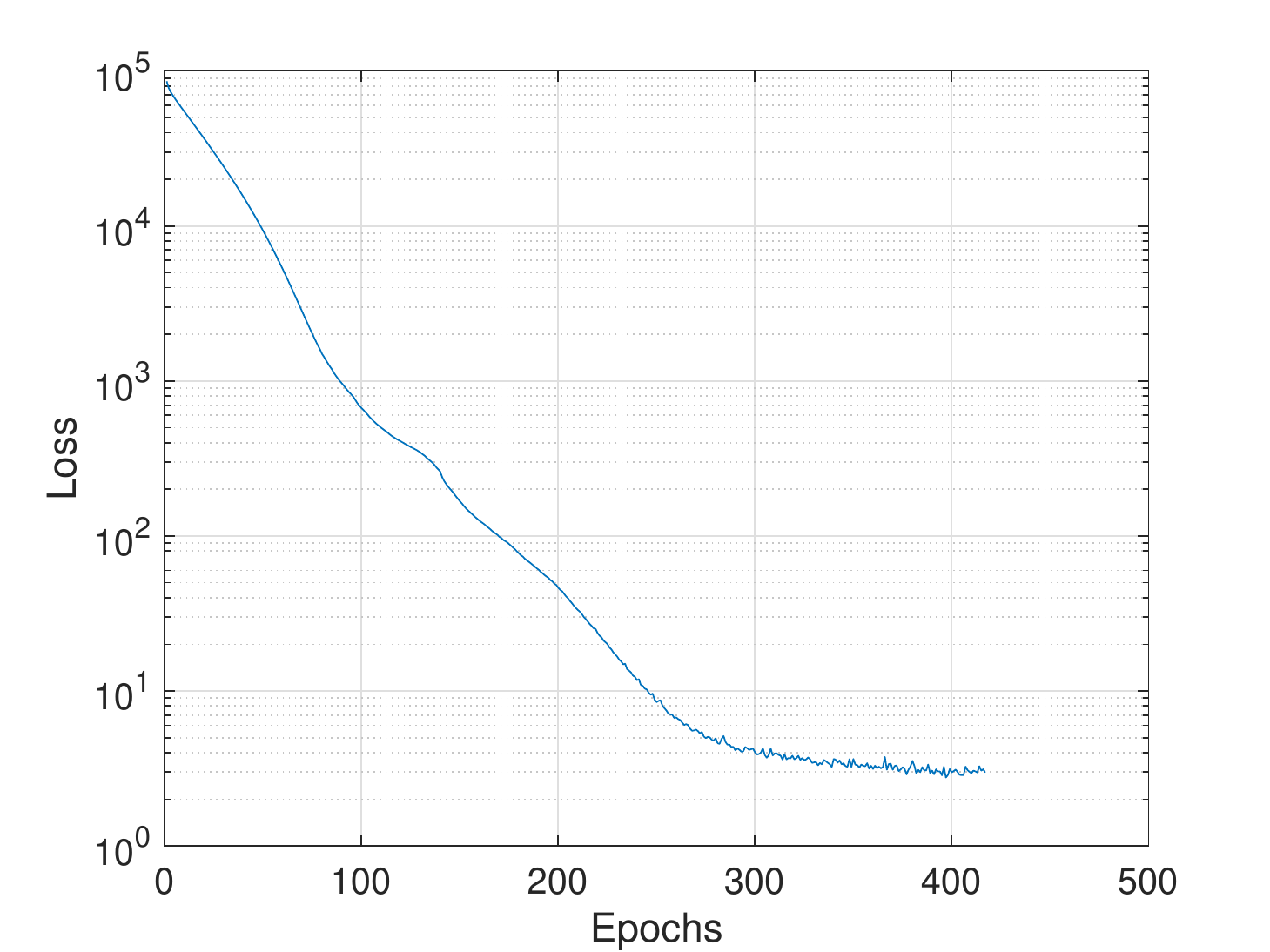}
   \caption{Loss value as a function of epochs. Early stopping is done before reaching 500 maximum epochs.}
	\label{fig:loss}
\end{figure}


Finally, to show convergence of DL algorithm, we plot loss value as function of number of epochs in Fig. \ref{fig:loss}. The curve is plotted for 2 hidden layers, 3 beams from the serving cell, 2 beams from the neighboring cell and cell-specific training; but other cases exhibit similar behaviour. Note that y-axis is in logarithmic scale. As expected, loss decreases rapidly with epochs and then saturates. We have set maximum number of epochs to 500, but because of early stopping criterion, the experiment stops earlier when no appreciable decrease in loss is observed for several epochs.

\subsection{Decision-Tree Regression}
For a comparison purpose, we evaluate positioning accuracy for our setup using Decision-Tree Regressor. In our feature vector, the features are modular. The UEs are first subdivided based on cell IDs. Further divisions are done based on beam IDs, and finally RSRP values help to differentiate between UEs with the same beam and cell ID. Decision Tree Regression suits our problem very well.

\begin{figure}
\centering
  	\includegraphics[width=3.5in]{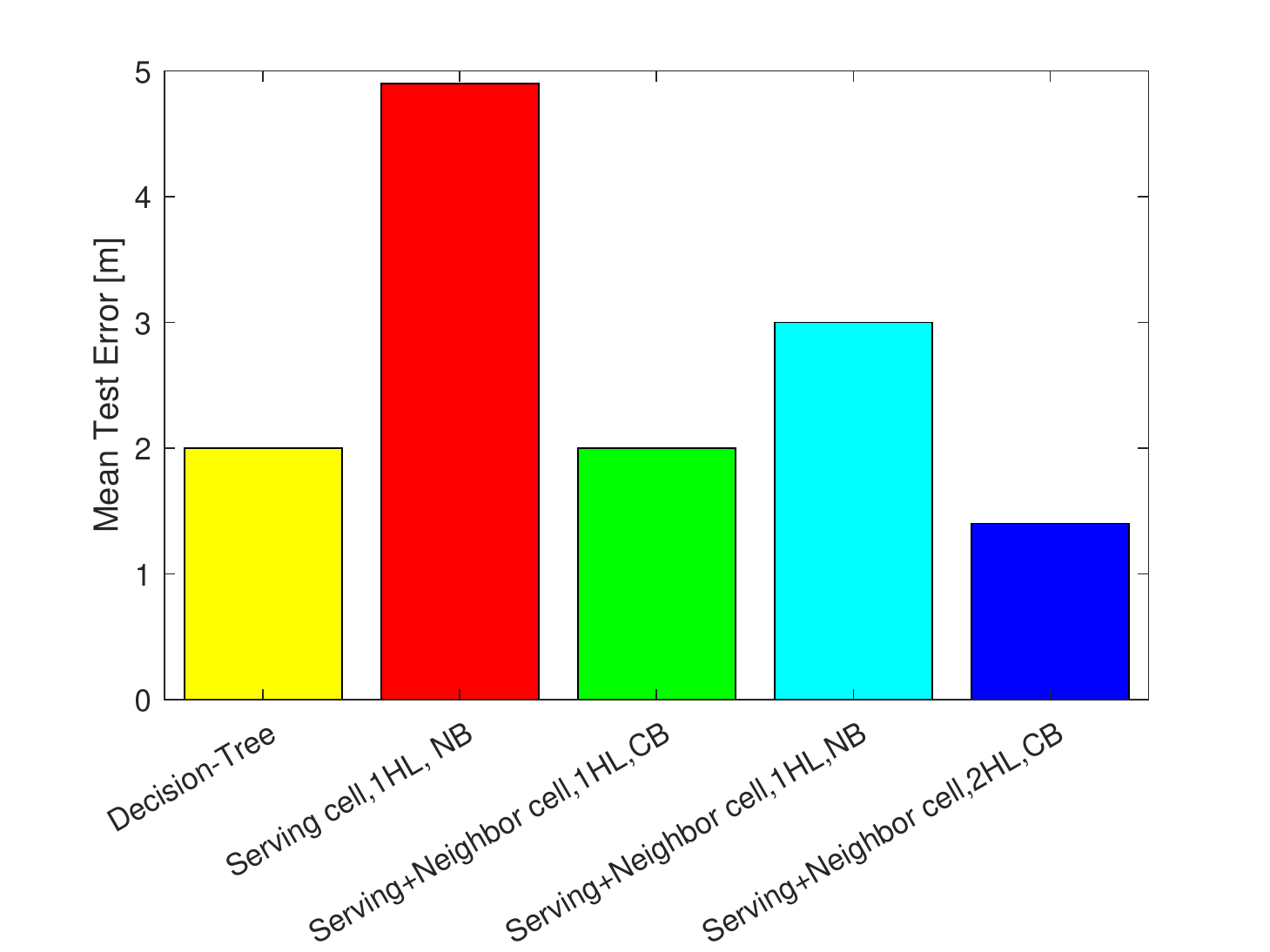}
   \caption{Performance comparison of Decision Tree Regressor and various DNN cases with different input features, number of hidden layers (HL) and architecture for training data, i.e. network based (NB) or cell-based (CB).}
	\label{fig:comp}
\end{figure}

We use Decision-Tree Regression approach for \emph{network level} training because it eliminates the need for cell-level training servers.
Fig. \ref{fig:comp} shows positioning accuracy for Decision tree approach. The best results are again obtained for the case with 3 beams from the serving cell and 2 from the neighboring cells. We get mean positioning error of 2.1m which is quite close but less than the mean positioning error obtained by DNN approach. We include a summary of results discussed in this paper and conclude that DNN with cell-based training using both serving as well as neighboring cell RSRP beam features provide the best results.

\section{Conclusion}
\label{sec:conclusions}
We study ML-assisted UE positioning in this work. We evaluate positioning accuracy for various combinations of feature vector based on beam RSRP and beam IDs from both serving and neighboring cells and search for the feature vector that provides smallest mean positioning error. Performance for both DNN and decision tree regression algorithms is studied. For a DNN, we studied network level and cell-specific DNN training models and quantified the performance. Cell-specific DNN with 2 layers provides mean positioning error of 1.4m by using beam RSRP reporting parameter alone as compared to conventional positioning algorithms that use angle and time of arrival parameters. This is even remarkable given that available raytracing data has minimum UE separation of one meter that serves as bench mark. As the nature of the feature vector resembles a decision tree, we also used a decision tree regression approach for comparison that provides closer but inferior result with mean positioning error of approximately 2m. Our proposed methodology to use raytracing, system level simulator and ML tools serves as a useful framework to evaluate potential effectiveness of deployment before derive testing and actual deployment.

\bibliographystyle{IEEEtran}
\bibliography{bibliography}
\end{document}